\documentclass{IOS-Book-Article}

\usepackage{mathptmx}

\usepackage{graphicx}
\usepackage{subcaption}
\usepackage{algorithm,algpseudocode} 
\usepackage{enumitem}	 
\usepackage{amsmath}	 
\usepackage{ltxtable}		
\usepackage{xcolor}		 
\usepackage{bm}				
\usepackage{amssymb}
\usepackage{float}			
\newcommand{\eatme}[1]{ }	
\usepackage[switch]{lineno} 	
\usepackage{tabularx} 
\usepackage{multirow}
\usepackage{blindtext}
\usepackage{booktabs}
\usepackage{hyperref}

\begin{document}
\begin{frontmatter}              

\title{Real-time Intersection Optimization for Signal Phasing, Timing, and Automated Vehicles' Trajectories}
\runningtitle{IOS Press Style Sample}

\author[A]{\fnms{Mahmoud Pourmehrab}%
\thanks{Corresponding Author: Mahmoud Pourmehrab, E-mails: mpourmehrab@ufl.edu}},
\author[A]{\fnms{Lily Elefteriadou}}
and
\author[B]{\fnms{Sanjay Ranka}}

\address[A]{Department of Civil Engineering, University of Florida, Gainesville, FL 32611, United States}
\address[B]{Department of Computer and Information Science and Engineering, University of Florida, Gainesville, FL 32611, United States}

\begin{abstract}
This study aims to develop a real-time intersection optimization (RIO) control algorithm to efficiently serve traffic of Connected and Automated Vehicles (CAVs) and conventional vehicles (CNVs).
This paper extends previous work to consider demand over capacity conditions and trajectory deviations by re-optimizing decisions.
To jointly optimize Signal Phase and Timing (SPaT) and departure time of CAVs, we formulated a joint optimization model which is reduced to and solved as a Minimum Cost Flow (MCF) problem.
The MCF-based optimization models is embedded into the RIO algorithm to operate the signal controller and to plan the movement of CAVs.
Simulation experiments showed 18-22\% travel time decrease and up to 12\% capacity improvement compared to the base scenario.
\end{abstract}

\end{frontmatter}

\thispagestyle{empty}
\pagestyle{empty}

\section{Introduction}
Autonomous\textemdash also called self-driving, driver-less, or unmanned\textemdash vehicles are equipped with hardware and software that automate the driving task.
Connected Vehicles (CVs) are those that can establish a two-way communication to other units for a variety of uses, ranging from synchronizing the relative movement of two vehicles to optimizing performance of a network of intersections.
Connected and Automated Vehicles (CAVs) \emph{combine} automation and connectivity by creating a reliable, efficient, and cooperative system to navigate the roads (\cite{Hobert2016}).

The CAV technology pursues mobility as the prime goal in two major ways.
First, the availability of high-resolution real-time data unlocks the potential to design traffic control systems which are demand-responsive at a micro-level.
A Vehicle-to-Infrastructure (V2I) web can collect real-time spatial data through Dedicated Short Range Communications (DSRC) or cellular network (4G, 5G).
The data can be used to optimize the performance of the traffic control systems in the form of active traffic management solutions to ease congestion and improve safety.
Among many recent proposed such applications are emergency vehicle preemption, queue warning, signal coordination, active pedestrian detection, and platoon promoting systems.
Second, AVs act as programmable units which can be instructed to drive in a system-wide efficient way.
Recent studies suggest various trajectory-based Intersection Management Algorithms (IMAs) to minimize the rate of accidents, travel time delay, and fuel consumption (\cite{7562449}).\par
The objective of this study is to develop an optimization-based intersection control algorithm that derives the Signal Phase and Timing (SPaT) and optimal CAVs trajectory functions for a mixed-traffic of CAVs and conventional vehicles (CNVs).
The proposed method extends the intersection control algorithm in \cite{PourmehrabarXiv} by developing real-time optimization methods for conditions in which re-optimization is necessary to account for demand exceeding capacity.
We formulated a unique mathematical model that optimizes the SPaT and CAVs departure times.
The model is reduced to a minimum cost flow network optimization where there exist efficient algorithms to obtain its global optimal solution at a complexity level as low as a linear program.   
Our approach also addresses changes in the anticipated trajectories of CNVs.
A constrained optimization model is developed to further minimize the space-time that a CAV consumes to reach the stop bar at the decided departure time.
In order to sustain optimality through time, the proposed framework allows for frequent re-optimization to make decisions adaptive to vehicles movement in real-time.
The framework is scalable for arbitrary intersection configuration, traffic, and CAV penetration rate.\par
The rest of the paper is organized as follows: Section \ref{lit_rev} reviews closely related works in the area of integrating CAV technologies with isolated signalized intersections.
Section \ref{methodo} develops the modules to the proposed algorithm.
In section~\ref{sec:phaseGen}, we define a set of feasible phases at a signalized intersection.
Section~\ref{sec:earliest} introduces the concept of \emph{earliest departure time} for incoming vehicles.
Section~\ref{sec:SPaT} defines the signal phase and timing (SPaT) problem followed by the development of an algorithm based on the minimum cost flow problem to solve it.
Section~\ref{sec:trjModels} explains the trajectory planning models for CAVs and CNVs.
Section \ref{sec:simExp} describes the scenarios and the simulation experiments followed by the numerical results.
Section \ref{sec:conclusion} summarizes the findings and provides recommendations for future research.\par

\section{Literature Review }\label{lit_rev}

This section provides an overview of some of the recent studies on integration of CAV technologies to enhance operation of isolated signalized intersections.
In the first category of algorithms, i.e. the reservation-based models, the control decision is to optimally schedule departure of vehicles through the intersection.
The second category, i.e. the trajectory-based, also optimizes the movement of CAVs to further coordinate SPaT with the traffic flow.
The following sections describe these two types of control algorithms.
A more comprehensive survey on control algorithms with communicative vehicles can be found in \cite{Li2014a,Florin2015,Chen2016,7562449}
and on AV perception technologies in \cite{VanBrummelen2018}.\par

\subsection{Reservation-based Signal Control Algorithms}
\cite{Dresner2004} first proposed an agent-based intersection control algorithm, called autonomous intersection manager (AIM), which serves AVs based on making reservations.
In their model, a central computer coordinates the movement of AVs and the traffic lights are eliminated.
The AIM divides an intersection conflict zone into reservable tiles and books them on a first-in-first-serve basis to no more than one AV.
In that study, they assumed no turning movements are available and vehicles drive at the speed limit.
Their results indicate 200-300\% reduction in delay compared to pre-timed control.
In continuation of their work, \cite{Dresner2008} extended AIM to allow for CNVs in the traffic stream.
They introduced a traffic-light model to periodically assign green time to CNVs in each approach.
The experiments showed almost no delay for scenarios with a full AV penetration rate.
However, for those scenarios with lower than 90\% AV ratios, delay increases exponentially as demand increases.\par
\cite{wu2012} proposed a heuristic control algorithm to determine the optimal departure sequence of AVs arriving at an intersection.
They showed the proposed heuristic-based AIM consistently decreases delay by 10-15\% when compared to adaptive control, while the reservation-based AIM proposed in \cite{Dresner2008} yields a relatively high delay for over-saturated conditions.
\cite{ahmane2013modeling} proposed a Timed Petri Nets with Multipliers (TPNM) model to obtain optimal departure sequence of AVs.
In their study, they argue the optimal sequence, without specified departure times, are more practical if AVs fail to follow a planned reservation.
\cite{sharon2017protocol} recently developed the Hybrid-AIM (H-AIM) on top of the AIM proposed in \cite{Dresner2008}.
H-AIM reserves space and time of the intersection for both AVs and CNVs.
With a predefined phase sequence, arrival data of the CNVs determine the green duration of a phase.
In their proposed method, AVs are allowed to run a red light if the requested intersection tiles are available.
Their simulation experiments show delay reductions at AV penetration rate of 90\% or higher.\par

\subsection{Trajectory-based Signal Control Algorithms}

\cite{li2006cooperative} proposed an intersection control algorithm which plans trajectories for a fully automated environment.
Their algorithm partitions incoming traffic into groups of three, based on their distance to the intersection.
They developed four trajectory planning algorithms which enumerate collision-free paths and sequence departure within the groups.
\cite{lee2012} developed an intersection control algorithm that minimizes the total overlap length of trajectories for traffic of AVs.
They observed 99\% delay reduction compared to actuated control in simulation.\par
\cite{li2014signal} proposed a joint signal and AV trajectory optimization model with a rolling horizon scheme.
In their work, a multi-component trajectory model with acceleration, deceleration, and constant speed components is used to optimize AVs movement.
They showed 16-37\% delay reduction for a single-lane through movement only intersection when compared to actuated control.
\cite{Yang2016} solved a similar problem for mixed traffic of CAVs and CNVs.
They kept the simplifying assumption of single-lane through movement only intersection and approximated the trajectory of CAVs as a piece-wise linear function.
They identified their method to be effective, compared to actuated control, only at 50\% CAV penetration rate or above.
\cite{zhuofei2018} extended the proposed algorithm in \cite{li2014signal} to optimize departure sequence and trajectory of AVs at an intersection with full turning movements.
The results improved delay by 16-79\% compared to actuated signal control.
\cite{FENG2018364} recently developed an algorithm to minimize fuel consumption at an isolated intersection with 100\% CAVs.
Their stage-wise model optimizes signal control using a Dynamic Programming paradigm, and the trajectory of vehicles using optimal control theory.
The method showed 24\% and 14\% delay reduction compared to pre-timed and adaptive controls, respectively.\par

\subsection{Summary}

This section outlined some of the intersection control algorithms in reservation- and trajectory-based categories.
For a full list of reviewed studies see Table~\ref{tab:litRevTable}.
The following identifies some of the gaps in the previously developed algorithms that this study addresses:
\begin{itemize}
    \item
    Previous studies do not address challenges with algorithm design for demand over capacity conditions.
    Over-saturated conditions result in vehicles having to wait for multiple phase switches before being served.
    Therefore, a more sophisticated optimization model, compared to under-saturated conditions, is needed to monitor vehicles over several phase changes.
    In over-saturated conditions there is a higher number of possible SPaT combinations and actual vehicles' performance is more likely to deviate from the one expected by the control algorithm.
    
    \item
    Most proposed methods serve incoming traffic through limited phasing schemes or in sequences with phasing restrictions (Table~\ref{tab:litRevTable}, second and fourth columns).
    An optimization model to link demand to SPaT decisions can enhance the adaptiveness through flexible selection and timing of phases.   
    \item
    Most developed algorithms only consider traffic consisting of full CAVs (refer to Table \ref{tab:litRevTable}).
    Several studies which incorporated mixed-traffic into the algorithm design did not clearly describe the difference in the trajectory of each vehicle type (Table~\ref{tab:litRevTable}, sixth and seventh columns).
    Hence, the trajectory planning model which is able to capture the effect of CNVs and CAVs to model their interaction in the traffic requires further research.
    
    
    
    
    
\end{itemize}

To address the gaps, we propose an algorithmic method that decides SPaTs and CAVs' trajectories based on a minimum cost flow network optimization.
    
\begin{table*}[htbp]
	\caption{Literature review summary table: development of control algorithms to operate an intersection with CAVs.}\label{tab:litRevTable}

\tiny
\begin{tabular} {p{0.5cm}p{1cm}p{.8cm}p{.5cm}p{.5cm}p{.8cm}p{.8cm}p{.8cm}p{1cm}p{2cm}}
	\toprule
	Study                          & Congestion effect modelling & Signalization                                                          & Traffic Composition & Traffic Light & Available movements     & Trajectory Planning & Car-following                                            & Simulation Platform & Results                                                                                                    \\
	\midrule                                                                                                                         
	\cite{Dresner2004}             & no                          & individual reservations                                                & AV                  & not present   & only throughs           & no                  &                                                          & Java                & 200-300\% reduction in avg delay compared to pre-timed control                                             \\
	\cite{wu2007discrete}          & no                          & three phases                                                           & AV                  & present       & only throughs           & no                  &                                                          & VISSIM              & 90\% decrease in waiting time compared to pre-timed control                                                \\
	\cite{li2006cooperative}       & no                          & optimal departure sequence in groups of three                          & AV                  & not present   & only throughs           & yes                 &                                                          &                     &                                                                                                            \\
	\cite{Dresner2008}             & no                          & individual reservations for AVs; Preodic Green Approach for CNVs       & CNV+AV              & present       & all                     & no                  &                                                          & Java                & Exponential delay with flows of less than 90\% AVs; Zero delay for traffic of 100\% AVs regardless of flow \\
	\cite{yan2008}                 & no                          & individual reservations                                                & AV                  & present       & only throughs           & no                  &                                                          &                     &                                                                                                            \\
	\cite{yan2009}                 & no                          & individual reservations                                                & AV                  & not present   & all                     & no                  &                                                          &                     & NA                                                                                                         \\
	\cite{zohdy2012}               & no                          & individual reservations                                                & AV                  & not present   & only throughs           & no                  &                                                          &                     & 70\% reduction in delay compared to stop control                                                           \\
	\cite{wu2012}                  & no                          & individual reservations                                                & AV                  & not present   & only throughs           & no                  &                                                          &                     & 10-15\% delay reduction compared to adaptive control                                                       \\
	\cite{lee2012}                 & no                          & NEMA phase set                                                         & AV                  & not present   & all                     & yes                 & minimized overlaped length of trajectory                 & VISSIM              & 99\% delay reduction compared to actuated control                                                          \\
	\cite{agbolosu2012quantifying} & no                          & NEMA phase set                                                         & CNV                 & present       & all                     & no                  &                                                          & VISSIM              & 12.5\% delay reduction compared to actuated control                                                        \\
	\cite{Xie2012}                 & no                          & optimal SPaT                                                           & CNV                 & present       & all                     & no                  & Krauss                                                   & SUMO                & 7-13\% avg. travel time reduction compared to pre-timed control                                            \\
	\cite{lee2013}                 & no                          & NEMA phase set                                                         & CV                  & present       & Throughs and left turns & no                  &                                                          & VISSIM              & 34\% delay reduction compared to actuated control                                                          \\
	\cite{ahmane2013modeling}      & no                          & optimal departure sequence                                             & AV                  & not present   & all                     & no                  &                                              &                     & 14-63\% avg stop time reduction compared to pre-timed control                                              \\
	\cite{li2014signal}            & no                          & optimal SPaT                                                           & AV                  & present       & only throughs           & yes                 & case-based multi-component trajectories                  & MATLAB              & 16-37\% delay reduction compared to actuated control                                                       \\
	\cite{Yang2016}                & no                          & optimal departure sequence                                             & CNV+CAV             & not present   & only throughs           & yes                 & Intelligent Driver Model                                 & Java                & 50\% penetration rate of CAVs required to outperform actuated control                                      \\
	\cite{Tachet2016}              & no                          & individual reservations                                                & AV                  & not present   & only throughs           & no                  & fundamental motion equations                             &                     & 83-97\% delay reduction compared to pre-timed control                                                      \\
\bottomrule
\end{tabular}
\end{table*}

\begin{table*}[htbp]
\begin{flushleft}
\textbf{Table \ref{tab:litRevTable}.} (continued).
\end{flushleft}

\tiny
\begin{tabular} {p{0.5cm}p{1cm}p{.8cm}p{.5cm}p{.5cm}p{.8cm}p{.8cm}p{.8cm}p{1cm}p{2cm}}
	\toprule
	Study                          & Congestion effect modelling & Signalization                                                          & Traffic Composition & Traffic Light & Available movements     & Trajectory Planning & Car-following                                            & Simulation Platform & Results                                                                                                    \\
	\midrule                                                                                                                         
	\cite{WEI2017102}              & no                          & pre-timed                                                              & CAV                 & present       & only throughs           & yes                 & Newell                                                   & GAMS                &                                                                                                            \\
	\cite{sharon2017protocol}      & no                          & individual reservations for AVs; fixed-time variable-sequence for CNVs & CNV+AV              & present       & all                     & no                  &                                                          & Java                & No or little improvement for 90\% or lower penetration rate of AVs                                         \\
	\cite{LI2017479}               & no                          & Individual reservation for AVs; fixed sequence for CNVs                & CNV+AV              & present       & all                     & no                  &                                                          & GAMS                & AVs reservation requests get canceled at 10\% or higher penetration rate of them                           \\
	\cite{Jiang2017}               & no                          & pre-timed SPaT                                                         & CNV+CV              & present       & only throughs           & yes                 & Intelligent Driver Model                                 & VISSIM              & 2-58\% reduction in fuel consumption compared to all CNV traffic                                           \\
	\cite{doi:10.1137/15M1048197}  & no                          & optimal SPaT                                                           & CNV                 & present       & all                     & no                  &                                                          & Ipop                &                                                                                                            \\
	\cite{PourmehrabarXiv}         & no                          & enhanced adaptive SPaT                                                 & CNV+AV              & present       & all                     & yes                 & Gipps for CNVs; Travel time/headway minimization for AVs & MATLAB              & 38-52\% avg travel time reduction compared to actuated control                                             \\
	\cite{YAO2018456}              & no                          & pre-timed SPaT                                                         & CNV+AV              & present       & all                     & yes                 & Gipps                                                    & MATLAB              &                                                                                                            \\
	\cite{zhuofei2018}             & no                          & optimal departure sequence                                             & AV                  & not present   & all                     & yes                 & case-based multi-component trajectories                  & Java                & 16-79\%  delay reduction compared to actuated control                                                      \\
	\cite{FENG2018364}             & no                          & NEMA phase set                                                         & CAV                 & present       & all                     & yes                 & Next Generation Simulation (NGSIM)                       & MATLAB              & 24\% and 14\% delay reduction compared to pre-timed and adaptive controls                                  \\
	\cite{YU201889}                & no                          & NEMA phase set                                                         & CAV                 & present       & all                     & yes                 & Newell                                                   &                     & 41-83\% delay reduction compared to actuated control                                                       \\                                                       
	\cite{GUO201954}               & no                          & four phases                                                            & CNV+CAV             & present       & all                     & yes                 & Shooting Heuristic                                       & VISSIM              & 36\% avg. travel time reduction compared to adaptive signal control                                        \\                                                       
	\bottomrule
\end{tabular}
\end{table*}


\newpage
\section{Methodology Overview}\label{methodo}
This section develops an algorithm that operates signals and provides CAVs with trajectories for over-saturated traffic of CAVs and CNVs.

Table \ref{tab:notationTrack} lists the sets, indices, parameters, variables, and functions that are used in this paper.\par
\begin{table}[htbp]
    \caption{Nomenclature}\label{tab:notationTrack}
    \centering
    \begin{tabularx}{\linewidth} {cX}
        \toprule
        \textbf{Symbol}                     & \textbf{Definition}                                                                                                                                                    \\
        \midrule
        \textbf{Sets}                       &                                                                                                                                                                        \\
        $L$                                 & set of incoming lanes at the intersection                                                                                                                              \\
        $C(l)$                              & the set of lanes which are conflicting with lane $l \in L$                                                                                                             \\
        $\Phi$                              & set of phases where a phase is a group of lanes with non-conflicting movements                                                                                         \\
        \addlinespace
        \textbf{Indices}                    &                                                                                                                                                                        \\
        $l \in L$                           & an incoming lane index                                                                                                                                                 \\
        $\phi \in \Phi$                     & a phase index                                                                                                                                                        \\
        $l_i$                               & indexes the i$th$ vehicle, $l_i\in \{1,\ldots,N_l\}$ where $N_l$ is the number of vehicles in lane $l \in L$                                                           \\
        \addlinespace
        \textbf{Parameters}                 &                                                                                                                                                                        \\
        $d_l$                               & demand in lane $l \in L$ (in $veh$)                                                                                                                            \\
        $p_\phi,p'_\phi$                               & pointers to phase $\phi \in \Phi$                                                                                                                            \\
        $l_l$                               & pointers to lane $l \in L$                                                                                                                            \\
        $\tau$                              & reaction time of drivers (only for CNVs)                                                                                                              \\
        $h$                                 & saturation headway (minimum time-headway between consecutive vehicles)                                                                                                 \\
        $\underline{g},\bar{g}$             & minimum and maximum green duration (in $seconds$), accordingly.                                                                                                          \\
        $y$                                 & yellow time duration (in $seconds$)                                                                                                                                      \\
        $ar$                                & all-red clearance duration (in $seconds$)                                                                                                                                \\
        $\Pi$                               & the phase-lane incidence matrix $\Pi=[\eta_{l\phi} :\ \forall \ l \in L,~\forall \ \phi \in \Phi]$, where $\eta_{l\phi}$ is 1 if lane $l$ in phase $\phi$, 0 otherwise \\
        $t_{l_i}^{0}$                       & detection time for vehicle $l_i$ (in $seconds$)                                                                                                                          \\
        $d_{l_i}^{0}$                       & detection distance for vehicle $l_i$ (in $meters$)                                                                                                                       \\
        $v_{l_i}^{0}$                       & detection speed for vehicle $l_i$ (in $m/s$)                                                                                                                             \\
        $\bar{v}$                           & speed limit near intersection, in $m/s$                                                                                                                                \\
        $\underline{a}_{l_i},\bar{a}_{l_i}$ & acceleration/deceleration for vehicle $l_i$, in $m/s^2$                                                                                                                 \\
        $\underline{j}_{l_i},\bar{j}_{l_i}$ & minimum/maximum jerk for vehicle $l_i$, in $m/s^3$                                                                                                                 \\
        \addlinespace
        \textbf{Variables}                  &                                                                                                                                                                        \\
        $t_{l_i}$                           & arrival time at the stop bar for vehicle $l_i$ (in $seconds$)                                                                                                            \\
        $d_{l_i}$                           & distance to stop bar for vehicle $l_i$ (in $meters$) at time $t_{l_i}$, equals to the length of the queue behind the stop bar plus a safe gap (0 if no queue exists)     \\
        $(g_{\phi_i})_{i=1}^{j}$            & the green time sequence to assign $g_{\phi_i}$ seconds of green time to phase $\phi_i$ in the order of $i=1$ to $i=j$                                                  \\
        \addlinespace
        \textbf{Functions}                  &                                                                                                                                                                        \\
        $f_{l_i}(t)$                        & space-time function for vehicle $l_i$, i.e. distance to stop bar at time $t$, in $m$                                                                                   \\
        $v_{l_i}(t)$                        & speed at time $t$ for vehicle $l_i$, equals to $-df_{l_i}(t)/dt$, note $f_{l_i}(t)$ is continuous, in $m/s$                                                                 \\
        $a_{l_i}(t)$                        & acceleration rate at time $t$ for vehicle $l_i$, equals to $dv_{l_i}(t)/dt$, note $v_{l_i}(t)$ is continuous, in $m/s^2$                                                    \\
        $j_{l_i}(t)$                        & jerk rate at time $t$ for vehicle $l_i$, equals to $da_{l_i}(t)/dt$, note $a_{l_i}(t)$ is continuous, in $m/s^3$                                                    \\
        \bottomrule
    \end{tabularx}
\end{table}

\begin{figure}[htbp]
   \centering
     \includegraphics[width=.8\linewidth]{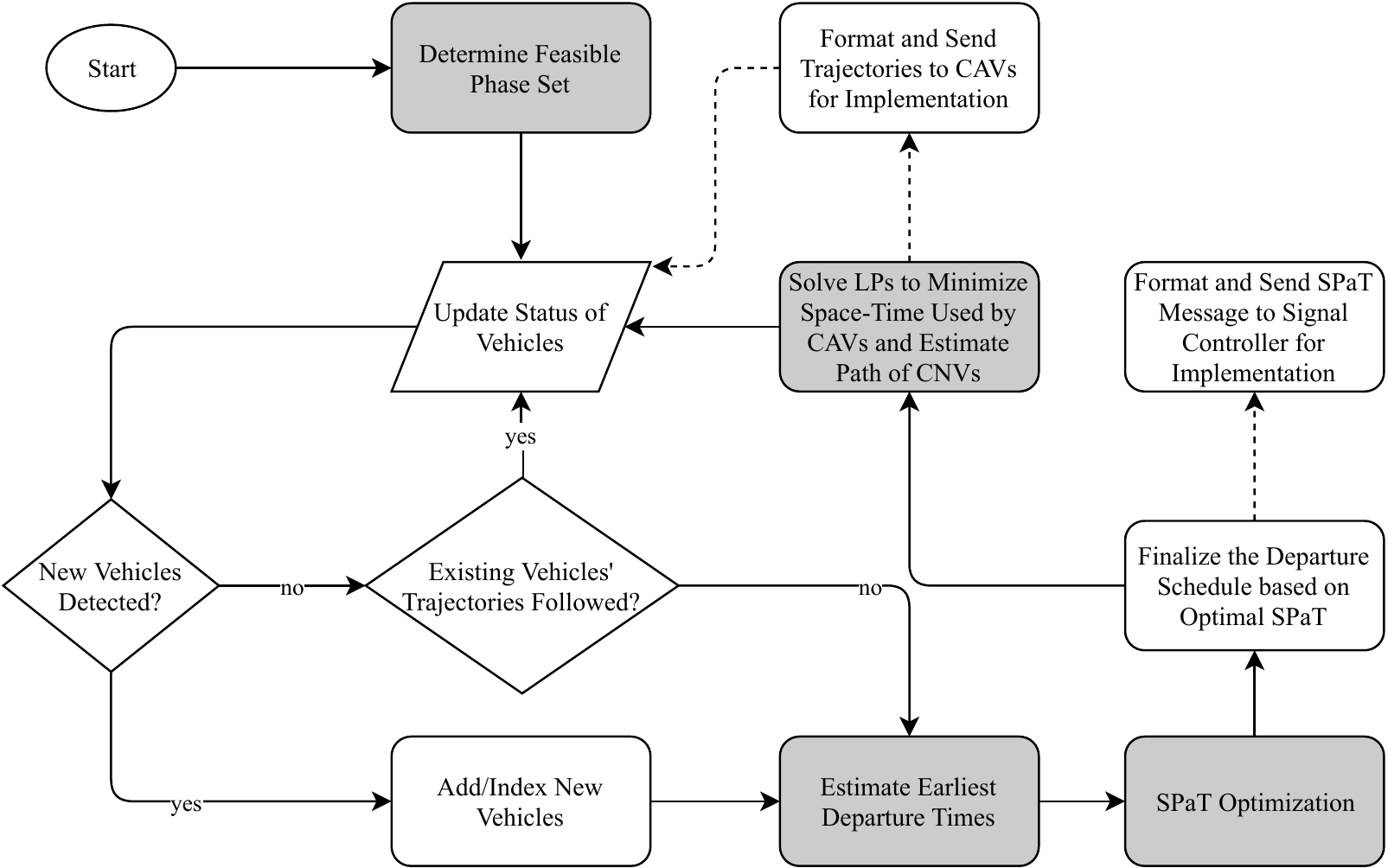}
   \caption{Real-time Intersection Optimization (RIO) Control Algorithm.}\label{fig:Overallflowchart}
 \end{figure}

Fig.~\ref{fig:Overallflowchart} provides the framework of the algorithm.
The process initializes with the determination of feasible phases at the intersection.
Next, the main control loop starts with updating vehicles' information.
Depending on the case, the update step may lead to either adding new vehicle data, adjusting fields of previously added vehicles, or both.  
If neither addition or adjustment operations are performed, the algorithm checks for the next status update.
Otherwise, a module estimates the earliest departure time of vehicles followed by SPaT optimization using the feasible set of phases.
The next step validates and finalizes the assigned departure times given the optimal SPaT sequence. 
At this stage the SPaT decisions and the actual departure time of vehicles are computed.
The information is used to optimize CAV trajectories to minimize used space-time driving toward the intersection.
The optimal SPaT decision and CAV trajectories are formatted and communicated to be implemented by the signal controller and the CAVs, respectively.
The control returns to the update stage and the main control loop repeats.\par

 
\begin{figure}[htbp]
\centering
\includegraphics[width=0.5\linewidth]{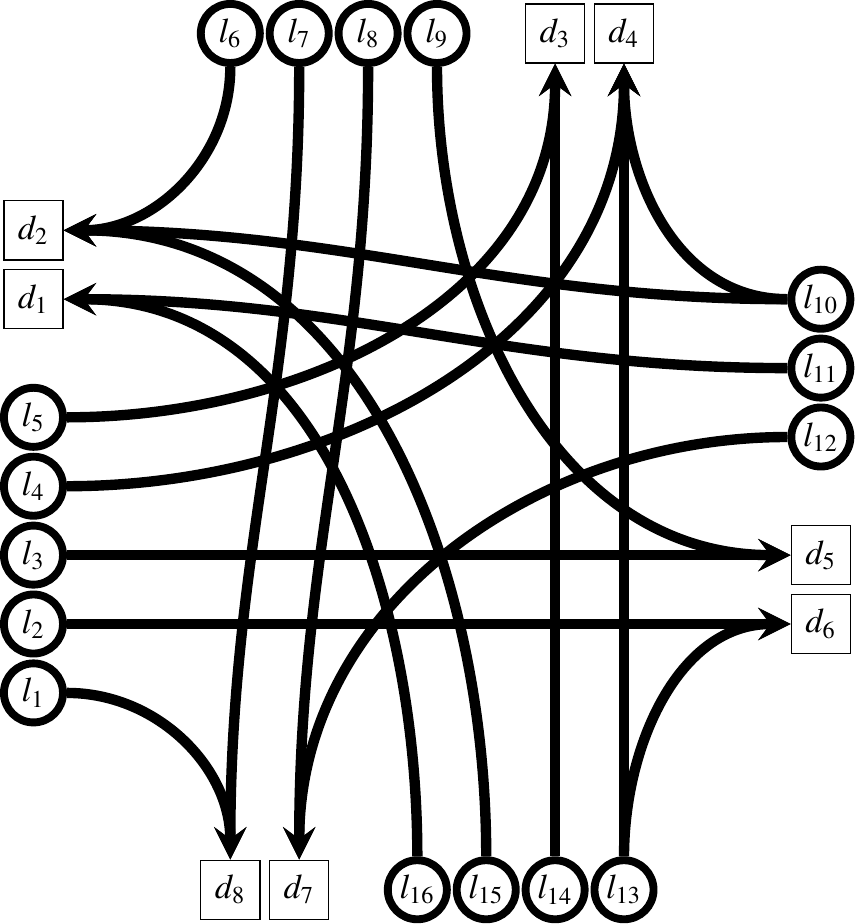}
    \caption{Channelization at the case study intersection.}\label{fig:inter_abs}
\end{figure}

The following subsections discuss the main parts of the algorithm.
We will use the real-world intersection of 13th and 16th (Gainesville, Florida) in Fig. \ref{fig:gMap} to illustrate the concepts introduced. The intersection has sixteen incoming lanes with the channelization shown in Fig. \ref{fig:inter_abs}.

\subsection{Determination of Feasible Phases}\label{sec:phaseGen}
We define a phase as a maximal inclusion-wise grouping of lanes with non-conflicting movements.
All the lanes within a phase share the same timing and color indication from the signal controller.
In order to avoid collisions, only the lanes in \emph{one} phase are allowed to receive a green indication while a red timing interval is allocated to other lanes.\par
Several previous studies assumed an intersection with two phases.
A typical four-leg intersection serves traffic through a more complex set of phases.
To maximize flexibility in determining the feasible phase set, an analyst can define any number of phases in the proposed RIO framework.\par

\begin{figure}[htbp]
\centering
\includegraphics[width=0.5\linewidth]{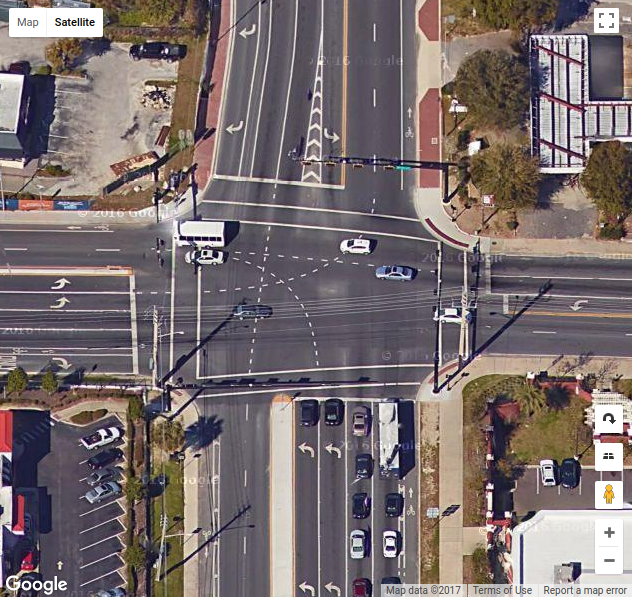}
\caption{Street view of the case study intersection (photo courtesy Google Maps).}\label{fig:gMap}
\end{figure}

\begin{figure}[htbp]
  \begin{center}
\includegraphics[width=0.9\linewidth]{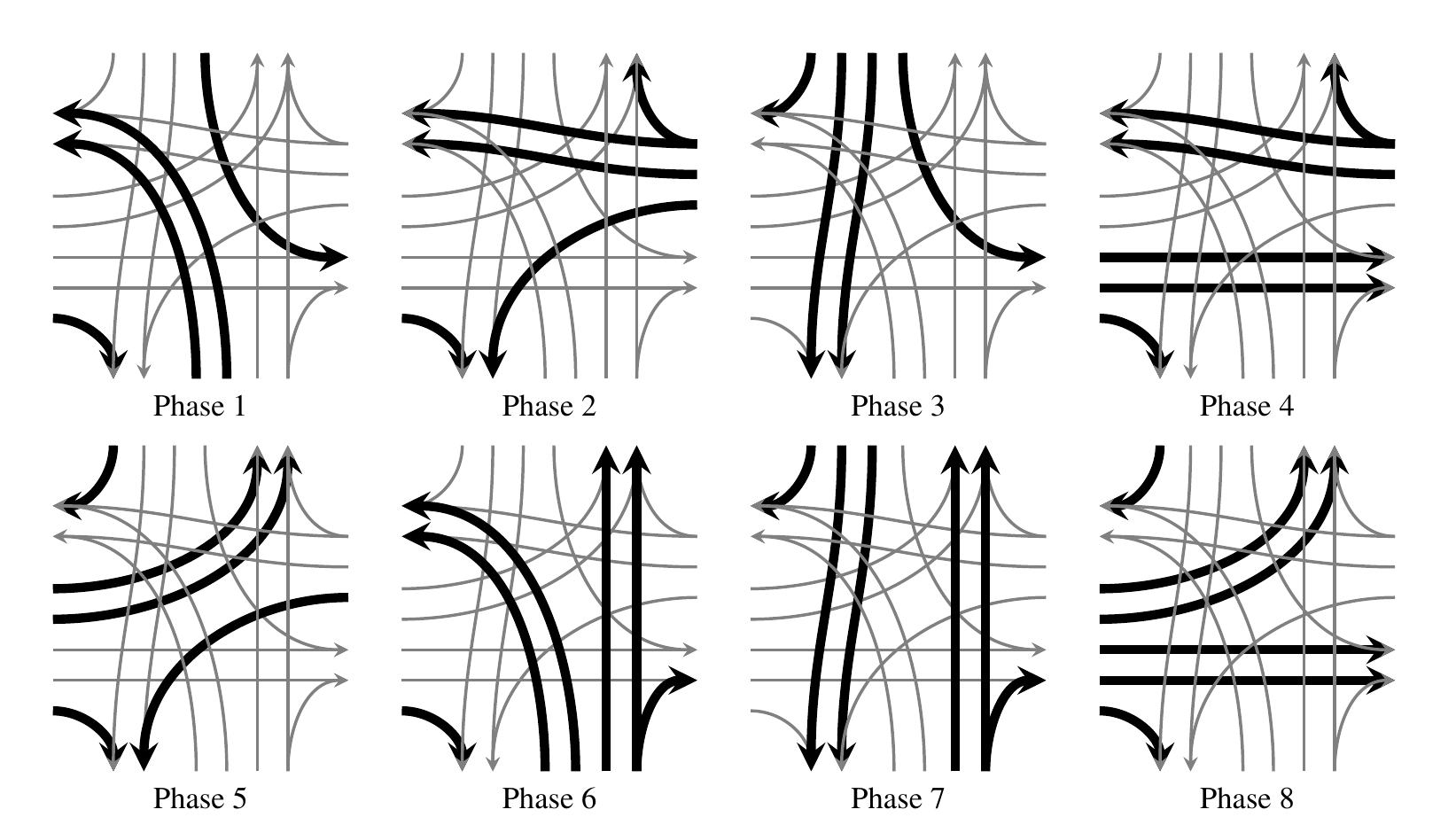}
    \caption{Feasible set of phases for case study intersection.}\label{fig:phases}
  \end{center}
\end{figure}

Fig. \ref{fig:phases} lists a collection of eight feasible phases for the case study intersection in Fig. \ref{fig:gMap}.\par


\subsection{Earliest Departure Time}\label{sec:earliest}
We define the term \emph{earliest departure time} for a vehicle to be the earliest time\textemdash if not influenced by the movement of other vehicles or red light\textemdash the vehicle will reach the stop bar.
When scheduling vehicles' departure time, knowledge of earliest departure time is used to prevent allocating unusable early green times which degrades intersection performance.\par

A variety of factors including vehicle type, initial speed, acceleration/deceleration capabilities, departure speed, and distance to stop bar can impact the earliest departure time.
For CNVs, the earliest departure time is estimated by considering the initial speed of each vehicle.
Under this assumption, at initial speed of $v_{l_i}$, a vehicle is not able to depart any earlier than $t_{l_i}^0+d_{l_i}/v_{l_i}$.\par
For CAVs, the earliest departure time can be estimated through estimation of a trajectory which minimizes the travel time of the vehicle.
It is noteworthy to mention the \emph{greedy} trajectory that grants the earliest departure time for a vehicle may be sub-optimal, if not infeasible, when considering all vehicles.
This is due to the fact that the optimal SPaTs seek to serve all vehicles at the least amount of time and through realistic trajectories.
In that sense, the earliest departure time must only serve as a lower bound on the actual departure time.\par

\subsection{Signal Phases and Timing}\label{sec:SPaT}
Consider an intersection with lanes $l \in L$ where the movements served in each lane are given.
Let sequence of triples $(g_{\phi_i},y_{\phi},ar_{\phi})_{i=1}^j$ to indicate green, yellow, and all-red duration for $j$ phases.
We define the SPaT problem to optimally select the phase and timing sequence to maximize \emph{green time utilization} by the vehicles\textemdash equivalent to maximizing throughput at the intersection level. A feasible solution to the stated SPaT problem must:
\begin{itemize}
    \item specify the number of phases to receive green, $j$,
    \item specify the phases and the order in which they should be executed, $(\phi_i)_{i=1}^j$,
    \item specify the amount of actual green assigned to each phase, $g_{\phi_i} \ \forall \ i=1, \ldots, j$,
    \item meet minimum and maximum green time restrictions, $\underline{g}\leq g_{\phi_i} \leq \bar{g}, \ \forall \ i=1, \ldots, j$,
    \item include only phase switches to a different phase, $\phi_i \neq \phi_{i+1} \ \forall \ i=1, \ldots, j-1$.
\end{itemize}
Hence, due to the mixed combinatorial nature of the decision space, achieving a global optimal SPaT is computationally costly.
Another challenge, besides optimality, is to devise a solution method that functions fast enough for real-time implementation.
Under simplifying assumptions, a variety of algorithms ranging from heuristic search methods to mathematical programming can be proposed.\par

Heuristic algorithms such as genetic algorithms, simulated annealing, particle swarm, and tabu search can optimize SPaT through a strategic random search.
The methods are less likely to be trapped in local optimal solutions since they can be designed to evenly sample the feasible regions.
However, for complex solution spaces, the searching process requires high CPU time which makes real-time optimization impractical.
To address this, typically, the run-time is controlled by limiting the number of iterations which, as a consequence, can lead to a suboptimal solution with low quality.\par

Mathematical programming, as an alternative to heuristic search, can be used to model and to solve the SPaT problem.
The challenge is to formulate an optimization problem that models vehicle interaction with SPaT at a low complexity.
In this study, we propose a model which is based on Minimum Cost Flow problem.\par

The minimum cost flow problem is a classic network optimization model to determine the cheapest flow pattern from the supply nodes to the demand nodes through capacitated arcs.
The problem can be formulated as a linear network model, if every arc has a constant unit cost.
As a result, efficient methods can be applied to solve a MCF model with constant cost functions.
The MCF model reads as:\par
\begin{align}
    & \min \label{eq:mcf_gen}
    \begin{aligned}[t]
    \sum_{\{(i,j) \in A\}} c_{ij} x_{ij}
    \end{aligned}  \\
                               & \text{subject to} \notag                                                                                             \\
                               & \sum_{\{j:(i,j) \in A\}} x_{ij}-\sum_{\{j:(j,i) \in A\}} x_{ji}=b(i) \quad \forall \ i \in N \label{eq:flow_bal_gen} \\
                               & l_{ij} \leq x_{ij} \leq u_{ij} \quad \forall \ (i,j) \in A \label{eq:range_gen}                                     
\end{align}
where,
\begin{itemize}
    \item[] $G=(N,A)$ the network composed of the set of nodes $N$ and the set of arcs $A$,
    \item[] $x_{ij}$ is the arc flow from node $i$ to node $j$,
    \item[] $c_{ij}$ is the arc unit cost from node $i$ to node $j$ (a constant),
    \item[] $l_{ij},u_{ij}$ are the minimum and maximum allowed arc flow from node $i$ to node $j$, respectively,
    \item[] $b(i)$ the net demand at node $i$,
\end{itemize}
For more details on minimum cost flow model specification and solution algorithms refer to \cite{ahuja2017network} chapters 9-11.

To solve the SPaT problem, we propose a network optimization model that can be reduced to an minimum cost flow (MCF) problem.
The proposed mathematical model aims to determine the unordered set of phases and their actual green times that best serve the detected number of vehicles in the lanes.  
The final algorithm that embeds the MCF-based model takes the earliest departure times, described in Section~\ref{sec:earliest}, and provides the optimal SPaT sequence and departure time of vehicles.\par

As shown in Table~\ref{tab:mcfNetTable}, we construct the proposed MCF network $G$ using four sets of arcs and their corresponding nodes. :
\begin{itemize}
    \item
    The first set of arcs distributes incoming vehicles in each lane, i.e. ($d_l \ \forall l \in L$, among phases which serve that lane.
    This is by supplying inflow of $d_l$ at the head of these arcs.
    Therefore, in the solution, the flow in arc $(l,p)$ represents the number of vehicles from lane $l$ which are considered to be served by phase $p$, if the maximum green time is enough to accommodate all.
    In order to incentivize serving the maximum non-conflicting lanes, the unit cost on these arcs is lower for those phases that serve higher number of lanes.
    The capacity of these arcs is the number of vehicles that can be served within the maximum green time.
    Based on the arc flows, we can determine the number of vehicles and the green time per phase.
    Hence, inactive phases \textemdash with zero inflow to their corresponding node in the network\textemdash are excluded from the sequence of phases and are not assigned any time.
    \item
    The third set of arcs $(l,r)$, can only have nonzero flow if a subset of vehicles could not be served due to maximum green limitation on the second set of arcs.
    In that case, the arc $(l,r)$ collects unserved vehicles in lane $l \in L$.
    \item
    Finally, the last set of arcs $(p,s)$ collects the flow corresponding to the vehicles to be served by phase $p \in \Phi$.
\end{itemize}

\begin{table}[htbp]
	\caption{Arcs for the proposed minimum cost flow model to solve the phase timing problem.}\label{tab:mcfNetTable}
	\centering
	\tiny
	\begin{tabularx}{\linewidth} {cccccc}
		\toprule
		Head           & Tail         & Cost                                    & Cap                         & Head Inflow & Tail Inflow             \\
		\midrule
		$l \in \phi_p$ & $p \in \Phi$ & $\max\limits_{p \in \Phi} \{ |p|\}-|p|$ & $\lfloor \bar{g}/h \rfloor$ & $d_l$       & $0$                     \\
		$l \in L$      & $r$          & $\infty$                                & $\infty$                    & $d_l$       & $-\sum_{l\in L} r_l$    \\
		$p \in \Phi$   & $s$          & $0$                                     & $\infty$                    & $0$         & $-\sum_{p\in \Phi} s_p$ \\
		\bottomrule
	\end{tabularx}
\end{table}
A simple flow balance on nodes indicates the number of served and unserved vehicles must be equal to the total number of incoming vehicles, or $\sum_{p\in \Phi} s_p+\sum_{l\in L} r_l=\sum_{l\in L} d_l$.\par

\begin{figure}[htbp]
  \begin{center}
\includegraphics[width=1\linewidth]{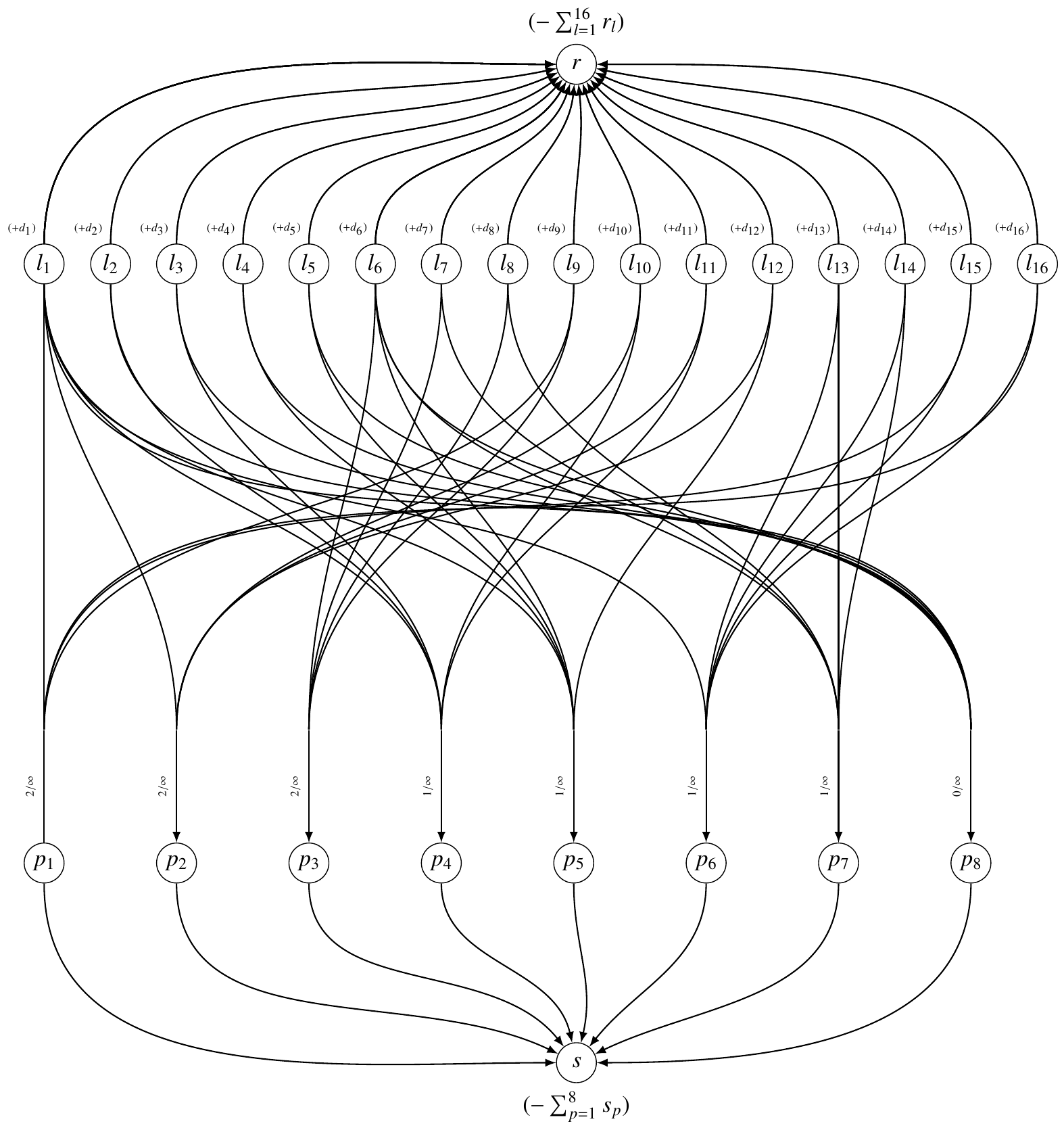}
    \caption{The phase time allocation problem for the test intersection. (The labels on the nodes indicate net demand. The labels on the arcs demonstrate cost/capacity, $0/\infty$ for arcs without label, as shown in table~\ref{tab:mcfNetTable}).}\label{fig:mcf}
  \end{center}
\end{figure}

For illustration purposes, Fig. \ref{fig:mcf} shows the phase time allocation network constructed for our study intersection shown in Fig. \ref{fig:inter_abs}.
The formulated phase time allocation problem reduces to a minimum cost flow problem with constant costs on its arcs.
Therefore, the problem is a linear network optimization and there exist several algorithms of polynomial complexity to solve it.
For the purpose of real-time implementation, we used \cite{cplex2018} API in Python programming language to formulate and solve instances of the problem within the proposed RIO framework.\par

Figure~\ref{fig:mcfAlgo} outlines the steps to making SPaT decisions based on the proposed network optimization model.
Once the MCF-based model is solved, the optimal values for flow in arcs $(p,p')$ shown as $x_p$ are available.
Assuming vehicles depart at saturation headway of $h$ \emph{seconds}, with the first vehicles delayed to react to green by $l_s$ \emph{seconds}, the green time can be computed as $g_p = max(\underline{g},(h-1)\times x_p+l_s) \ \forall p\in \Phi: x_p>0$.
At this point the optimal unordered phase collection and the allocated green times are obtained.
Next, the optimal phases are sorted on ascending order of the average earliest travel times for the served vehicles.
The optimal SPaT for the served vehicles is fully specified.
The same procedure is repeated only for the unserved vehicles up to the times that all incoming vehicles are served.\par

\begin{figure}[htbp]
    \begin{algorithmic}[1]
        \Require Number of vehicles in incoming lanes, Set of all phases
        \Ensure The optimal SPaT sequence $(g_{\phi_i})_{i=1}^j$
        \Procedure{MCF\_Optimizer}{}
        \State Solve MCF-based model \eqref{eq:mcf_gen}-\eqref{eq:range_gen}
        \State Set green times to $max(\underline{g},(h-1)\times x_p+l_s)$
        \State Sort phases based on ascending order of avg earliest departure time of served vehicles 
        \State Execute SPaT from the front of the sorted sequence
        \EndProcedure
    \end{algorithmic}
    \caption{The MCF-based SPaT Optimization Pseudo Code.}\label{fig:mcfAlgo}
\end{figure}

The advantage of using the proposed MCF-based algorithm is three-fold.
First, the model distributes demand of vehicles in lanes among phases.
The selected phases serve the highest number of vehicles by optimizing for use of phases with a higher number of lanes.
The duration and order of selected phases are determined based on demand in lanes to maximize green utilization at the intersection.
Second, the linear network optimization model is low in complexity level which makes it quick and reliable for frequent re-optimization in practice.
Third, considering the size and linearity of the model, the solution is guaranteed to be globally optimal.

\subsection{Trajectory Planning}\label{sec:trjModels}
The problem of trajectory planning aims to determine the longitudinal movement of a vehicle in an incoming lane to meet the departure schedule and speed given by the SPaT optimizer discussed in subsection~\ref{sec:SPaT}.
The vehicles' trajectories can be represented as a space-time relationship where the underlying association of time and space is case-dependent.
For instance, a follower vehicle's movement can be influenced by the trajectory of the vehicle ahead\textemdash i.e, the lead vehicle.
A lead vehicle, however, may drive under the impact of the traffic light facing the lane it is located at, independent of other vehicles.
Another behavioral distinction can be drawn based on the automation and the connectivity level of a vehicle.
CNVs' motion is estimated based on factors related to the driver, vehicle, and environment (see \cite{elefteriadou2016introduction}).
A CAV movement can be advised through an optimization process since those are equipped with on board-units to receive and follow the advisory information.\par

\subsubsection{Trajectory Model for CAVs}\label{sec:CAVtrjModels}
Let $f_{l_i} (t): [t_{l_i}^0, t_{l_i} ] \rightarrow [d_{l_i},d_{l_i}^{0}]$ to define the space-time relation that gives distance of the vehicle $l_i$ from the stop bar at any given time $t$.
We refer to $f_{l_i}(t)$ as trajectory of the vehicle that describes its movement from the initial location of $d_{l_i}^{det}$ from the stop bar (at time $ t_{l_i}^0$) to $d_{l_i}$ (at time $ t_{l_i}$).
The first and second derivatives of the trajectory, if $f(.)$ belongs to class $C^1$ continuity, yield the negative of vehicle's speed and acceleration rate at time $t$, respectively, i.e. $v_{l_i}(t)=-\frac{df_{l_i}(t)}{dt} ,\ a_{l_i}(t)=-\frac{df_{l_i}^2(t)}{dt^2} \ \forall \ t \in [t_{l_i}^0, t_{l_i}]$.
The rest of this subsection describes the set of constraints and objective function to specify the best trajectory function.\par
\paragraph{Control the Speed Profile}: A valid trajectory for a CAV prevents speeding and bounds the speed profile to be non-negative.
This can be represented as
\begin{subequations}
  \begin{align}
  0 \leq v_{l_i}(t) \leq \bar{v}.
  \end{align}
\end{subequations}  \par
\paragraph{Comply with Tracked Distance and Speed Information}: The trajectory should comply with the vehicle's provided data on speed and distance from the stop bar at the time stamp of the message.
The following set of constraints declares a feasible trajectory to be compatible with the distance and speed at the detection time-stamp:
\begin{subequations}
  \begin{align}
f_{l_i}(t_{l_i}^0) = d_{l_i}^0, 
v_{l_i}(t_{l_i}^0) = v_{l_i}^0.
  \end{align}
\end{subequations}
\paragraph{Comply with Scheduled Distance and Speed Information}: Optimization of SPaT decision determines the optimal departure schedule of vehicles at the maximum departure speed from the stop bar.
The following hard constraints guide the CAV to depart at the optimized departure time $t_{l_i}^0$ at maximum speed $v_{max}$, or mathematically:
\begin{subequations}
  \begin{align}
f_{l_i}(t_{l_i}^0) = 0, \\
v_{l_i}(t_{l_i}^0) = \bar{v}.
  \end{align}
\end{subequations}

\paragraph{Meet acceleration/deceleration and jerk limits}: A valid trajectory for a CAV has to keep acceleration/deceleration or jerk within a range that is comfortable for passengers and executable by the vehicle:
\begin{subequations}
  \begin{align}
  \underline{a}_{l_i} \leq a_{l_i}(t) \leq \bar{a}_{l_i}, \\
 \underline{j}_{l_i} \leq j_{l_i}(t) \leq \bar{j}_{l_i}.
  \end{align}
\end{subequations}

\paragraph{Objective Function}: While minimizing either occupancy or travel time of vehicles may lead to inefficient use of the the other resource, the area under the trajectory curve provides the ideal metric that combines the two factors.
Therefore, CAV trajectory optimizer minimizes the area under the trajectory curve, or mathematically:
\begin{subequations}
  \begin{align}
\min_{f_{l_i}, t_{l_i}} \int_{t_{l_i}^{0}}^{t_{l_i}} f_{l_i}(t).
  \end{align}
\end{subequations}\par
Finally, the model to minimize space-time used by a lead CAV reads as:
\begin{subequations}
  \begin{align}
    & \text{arg}\min_{f_{l_i}, t_{l_i}} \int_{t_{l_i}^{0}}^{t_{l_i}} f_{l_i}(t) \ dt \label{eq:leadCAV0}\\
    & \text{subject to} \notag \\
      & f_{l_i}(t_{l_i}^0) = d_{l_i}^0 ,                 & \label{eq:leadCAV1}                                       \\
      & v_{l_i}(t_{l_i}^0) = v_{l_i}^0 ,           & \label{eq:leadCAV2}                                       \\
      & f_{l_i}(t_{l_i}) =0 ,                      & \label{eq:leadCAV3}                                       \\
      & v_{l_i}(t_{l_i}) =\bar{v} ,                      & \label{eq:leadCAV3.5}                                       \\
      & 0 \leq v_{l_i}(t) \leq \bar{v}                   & \forall \ t \in (t_{l_i}^0, t_{l_i} ),\label{eq:leadCAV4} \\
      & \underline{a}_{l_i} \leq a_{l_i}(t) \leq \bar{a}_{l_i} & \forall \ t \in (t_{l_i}^0, t_{l_i} ).\label{eq:leadCAV5}\\
      & \underline{j}_{l_i} \leq j_{l_i}(t) \leq \bar{j}_{l_i} & \forall \ t \in (t_{l_i}^0, t_{l_i} ).\label{eq:leadCAV6}
  \end{align}
\end{subequations}
For a follower CAV, in addition to \eqref{eq:leadCAV1}-\eqref{eq:leadCAV6}, a set of constraints to keep a safe headway with the front vehicle, i.e,
\begin{align}
f_{l_i}(t) \geq f_{l_{i-1}}(t) + h \qquad \forall \ t \in (t_{l_i}^0, t_{l_i} ) \label{eq:CAVfolHeadway_before}
\end{align}
is necessary. \par
\begin{figure}[htbp]
  \begin{center}
    \includegraphics[width=.8\linewidth]{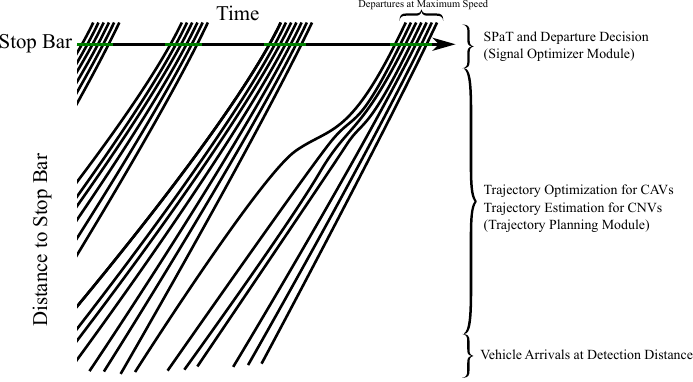}
    \caption{The trajectory optimization model ensures optimal SPaT schedule to depart vehicles at maximum speed and within the decided green times.}\label{fig:diff_k}
  \end{center}
\end{figure}

Before any knowledge of the functional form of space-time curve $f_{l_i}(t)$, the optimization model~(\ref{eq:leadCAV0}-\ref{eq:leadCAV5}) is undefined.
In this study, we make the following assumptions in order to parameterize the trajectory function $f_{l_i}(t)$:
\begin{itemize}
  \item The space-time relation can be approximated by a \emph{polynomial of degree} $k$, i.e. $f_{l_i}(t)=\sum_{n=0}^k \beta_{l_i,n} \times (\frac{t}{t_{l_i}-t_{l_i}^0})^n$.
  \item Speed and acceleration/deceleration rate are \emph{controlled} at $m$ uniformly distributed points within the time interval $(t_{l_i}^0, t_{l_i} )$. This transforms the domain of constraints (\ref{eq:leadCAV4},\ref{eq:leadCAV5},\ref{eq:CAVfolHeadway_before}) from a continuous interval to a set of points $M=1, \ldots,m$ (excluding the boundaries).
\end{itemize}
Under the above assumptions along with use of relative time measured from the detection time ($t_{l_i}^0=0$), the model~(\ref{eq:leadCAV0}-\ref{eq:leadCAV5}) simplifies to:
\begin{subequations}
	\begin{align}\label{eq:polyLead0}
		\textrm{LCAV: }& \text{arg}\min_{\beta_{l_i,n}}
		\begin{aligned}[t]
		\sum_{n=0}^k \frac{t_{l_i}}{n+1} \times \beta_{l_i,n}
		\end{aligned} \\
		&\text{subject to} \notag \\
		& \beta_{l_i,0} = d_{l_i}^0 , \label{eq:polyLeadC0}\\
		& \beta_{l_i,1} = - v_{l_i}^0 , \\
		& \sum_{n=0}^k \beta_{l_i,n} =0 , \\
		& \sum_{n=1}^k n \times \beta_{l_i,n} =-\bar{v} , \\
		  & 0 \leq \sum_{n=1}^k \frac{-n}{t_{l_i}} \times \bigg(\frac{j}{m+1}\bigg)^{n - 1} \times \beta_{l_i,n}  \leq \bar{v}                                        \notag\\& \forall \ j =1,\ldots,m,                     \\
		  & \underline{a}_{l_i} \leq \sum_{n=2}^k \frac{-n \times (n-1)}{t_{l_i}^2} \times \bigg(\frac{j}{m+1}\bigg)^{n - 2} \times \beta_{l_i,n}  \leq \bar{a}_{l_i} \notag\\& \forall \ j =1,\ldots,m.\label{eq:polyLead1} \\
		  & \underline{j}_{l_i} \leq \sum_{n=3}^k \frac{-n \times (n-1)\times (n-2)}{t_{l_i}^3} \times \bigg(\frac{j}{m+1}\bigg)^{n - 3} \notag\\&\times \beta_{l_i,n}  \leq \bar{j}_{l_i} \forall \ j =1,\ldots,m.\label{eq:polyJ} 
	\end{align}
\end{subequations}
Similarly, transformation of constraint \eqref{eq:CAVfolHeadway_before} provides the following inequality to assure safe time headway between the vehicles:
\begin{align}
  \sum_{n=0}^k \beta_{l_i,n} \times (\frac{j}{m+1})^n  \geq \sum_{n=0}^k \beta_{l_{i-1},n} \times (\frac{j}{m+1})^n  +  h  \notag\\\forall \ j =1,\ldots,m\label{eq:CAVfolHeadway_after}
\end{align}
Finally, the trajectory optimization model for follower CAVs reads as:
\begin{align}
    \textrm{FCAV: }& \text{arg}\min_{\beta_{l_i,n}} \label{eq:polyFol0}
  \begin{aligned}[t]
  \sum_{n=0}^k \frac{t_{l_i}}{n+1} \times \beta_{l_i,n}
  \end{aligned}  \\
    & \text{subject to} \notag                                                              \\
    & \text{Equations~(\ref{eq:polyLeadC0}-\ref{eq:polyLead1})~\text{and}~(\ref{eq:CAVfolHeadway_after}).} \notag
\end{align}
The higher degree of polynomial $k$ makes the trajectory function more flexible within the allowed ranges of speed and acceleration.
However, after a certain threshold in $k$, the trajectory function becomes strongly restricted by upper bound on speed/acceleration constraints and the area under the curve converges to a limit value.
Figure~\ref{fig:diff_k} demonstrates how the trajectory planning stage fills the gap between the detected arrivals and the optimized departure schedule through smooth trajectory functions.\par
Both models (\ref{eq:polyLead0}-\ref{eq:polyLead1}) and (\ref{eq:polyFol0}) are constrained mathematical programs which are linear in the coefficient vector that defines the polynomial, i.e, $[\beta_{l_i,n}]$.
Hence the Linear Programs (LPs) are solved using \cite{cplex2018} solver in real-time.
The efficiency in solving the LPs makes it possible to choose the degree of polynomial $k$ free of computational concerns.\par

\subsubsection{Trajectory Model for CNVs}\label{sec:CNVtrjModels}
Similar to CAVs, CNVs' longitudinal movement can be described by a trajectory function $f_{l_i}(t)$.
However, since they are non-communicative, their arrival data are collected using devices such as camera, radar, or loop detectors, unlike CAVs which send the data through DSRC and radios.
In this section, we assume a \emph{lead} CNV keeps constant speed when arriving at the communication range.
For follower CNVs, the Gipps Car-Following model \cite{Gipps1981} relates the speed of a follower conventional vehicle to spatial attributes of the vehicle in front of it through the following:

{\tiny
\begin{align}\label{eq:Gipps}
	&v_{n_l}(t+\Delta t) =  \min \Bigg\{  v_{n_l}(t)+2.5 a_{n_l}^{acc} \times \Delta t \times (1-\frac{v_{n_l}(t)}{V_{n_l}^{des}}) \times \sqrt{0.025+\frac{v_{n_l}(t)}{V_{n_l}^{des}}}, a_{n_l}^{dec} \times \Delta t+\notag \\
	  &\sqrt{a_{n_l}^{dec} \times \bigg(2\times \Big(d_{(n-1)_l}(t)-d_{n_l}(t)+L_{n_l} \Big)+\Delta t \times \big(a_{n_l}^{dec} \times \Delta t+v_{n_l}(t) \Big)+\frac{v_{(n-1)_l}(t)^2}{a_{n_l}^{dec}}} \bigg) 
	\Bigg\}
\end{align}
}%

where:
\begin{enumerate}\footnotesize
\item[]$\Delta t$ is the time steps to compute trajectory points and the reaction time of CNVs
\item[]$v_{n_l}(t+\Delta t)$ is the speed of follower vehicle $\Delta t$ seconds after $t$
\item[]$L_{n_l}$ is the length of $n$th vehicle in lane $l$
\end{enumerate}

Using the Gipps model, a conventional vehicle speed profile can be estimated within the distance range of interest from the center of the intersection.
Then assuming constant-acceleration Fig. \ref{fig:GippsAlgo} describes the algorithm to compute the full trajectory of the follower vehicle $l_i$.
The algorithm starts from constructing the trajectory from time $t_{l_i}^0$, when the vehicle is initially detected and uses Gipps equation~(\ref{eq:Gipps}) to compute the speed one step forward.
Next, it computes the acceleration rate over the associated time interval.
Simple use of the constant-acceleration motion equation gives the distance of vehicle to the stop bar, i.e, $f_{l_i}(t)$.\par

\begin{figure}[htbp]
        \begin{algorithmic}[1]
            \Require $\tau,\ v_{l_i}(t),\ v_{l_i}^{des},\ a_{l_i}^{acc},\ a_{l_i}^{dec}, \ d_{l_i}(t),\ L_{l_i},\ d_{l_{i-1}}(t) $
            \Ensure $f_{l_i}(t)$

            \Procedure{Gipps\_Estimator}{}
            \State $t \gets t_{l_i}^0$
            \While{$t+\tau \leq t_{l_{i-1}}$}
            \State{Obtain $v_{l_i}(t+\tau)$ using Equation~(\ref{eq:Gipps})}
            \State{$a_{l_i}(t^\prime) \gets \frac{v_{l_i}(t+\tau)-v_{l_i}(t)}{\tau} \quad \textnormal{for} \quad t^\prime \in [t,t+\tau]$}
            \State{$d_{l_i}(t^\prime) \gets d_{l_i}(t) - v_{l_i}(t) \times (t^\prime-t) - \frac{a_{l_i}(t^\prime)}{2} \times (t^\prime-t)^2  \quad \textnormal{for} \quad t^\prime \in [t,t+\tau]$}
            \State{$t \gets t+\tau$}
            \EndWhile

            \Return{$f_{l_i}(t)=d_{l_i}(t) \quad \forall \ t \in [t_{l_i}^0,t_{l_{i-1}}]$}
            \EndProcedure
        \end{algorithmic}
    \caption{The Gipps car-following Trajectory Estimator \cite{Gipps1981}.}\label{fig:GippsAlgo}
\end{figure}


\section{Simulation Experiments and Results}\label{sec:simExp}
This section applies the proposed RIO algorithm to optimize SPaT and CAVs movement at the test intersection shown in Fig.~\ref{fig:gMap}.
We model the channelization of the test intersection with 16 incoming lanes and full set of movements as shown in Fig.~\ref{fig:inter_abs}.

A set of scenarios with the following parameters were designed:\par
\begin{itemize}
    \item Four incoming flow levels ranging from under-saturated to over-saturated are considered with exponentially distributed inter-arrival times (Table~\ref{tab:scenarios}).
    This is to quantify effectiveness of the proposed optimization algorithm when traffic transitions to highly congested conditions.
    
    \item At any incoming flow, eleven CAV ratios ranging from 0 to 1 are examined (Table~\ref{tab:scenarios}).
    CNVs and CAVs are uniformly distributed within the traffic stream.
    
    \item A minimum green of 5 seconds, a maximum green of 30 seconds, a yellow duration of 3.5 seconds, an all-red duration of 2 seconds with the set of phases shown in Figure \ref{fig:phases} were set as signal parameters.
    Although a higher maximum green improves performance in high demand scenarios, the maximum green is kept fixed in order to compare numerous scenarios with a consistent set of input.
    
\end{itemize}

\begin{table}[htbp]\footnotesize
    \caption{Traffic scenarios based on incoming flow (volume) and CAV ratios.}
    \label{tab:scenarios}
    \centering
    \begin{tabular}{p{0.45\linewidth}p{0.45\linewidth}}
        \toprule
        \textbf{Volume (vphpl$^\dag$)} & \textbf{CAV ratios}                                  \\
        \midrule
        250                                                        & 0.0 to 1.0 (step size = 0.1) \\
        450                                                        & 0.0 to 1.0 (step size = 0.1) \\
        650                                                        & 0.0 to 1.0 (step size = 0.1) \\
        850                                                        & 0.0 to 1.0 (step size = 0.1) \\
        \bottomrule
        \addlinespace
        \multicolumn{2}{l}{{$^\dag$ \tiny vehicle per hour per lane.}}\\
    \end{tabular}
\end{table}

The algorithm is implemented in Python 3.7 programming language and was run on an Ubuntu machine with Intel Core i7-8550U CPU and 8 GB RAM with no noticeable delay per iteration.
The arrival time, departure time, and throughput data were collected at vehicle level during each run.
The frequency of updating the optimization is set to 5 $Hz$ and an interface to CPLEX solver to compute the optimal solution to the LP sub-problems was used.\par

\begin{figure}[htbp]
  \centering
    \includegraphics[width=0.98\linewidth]{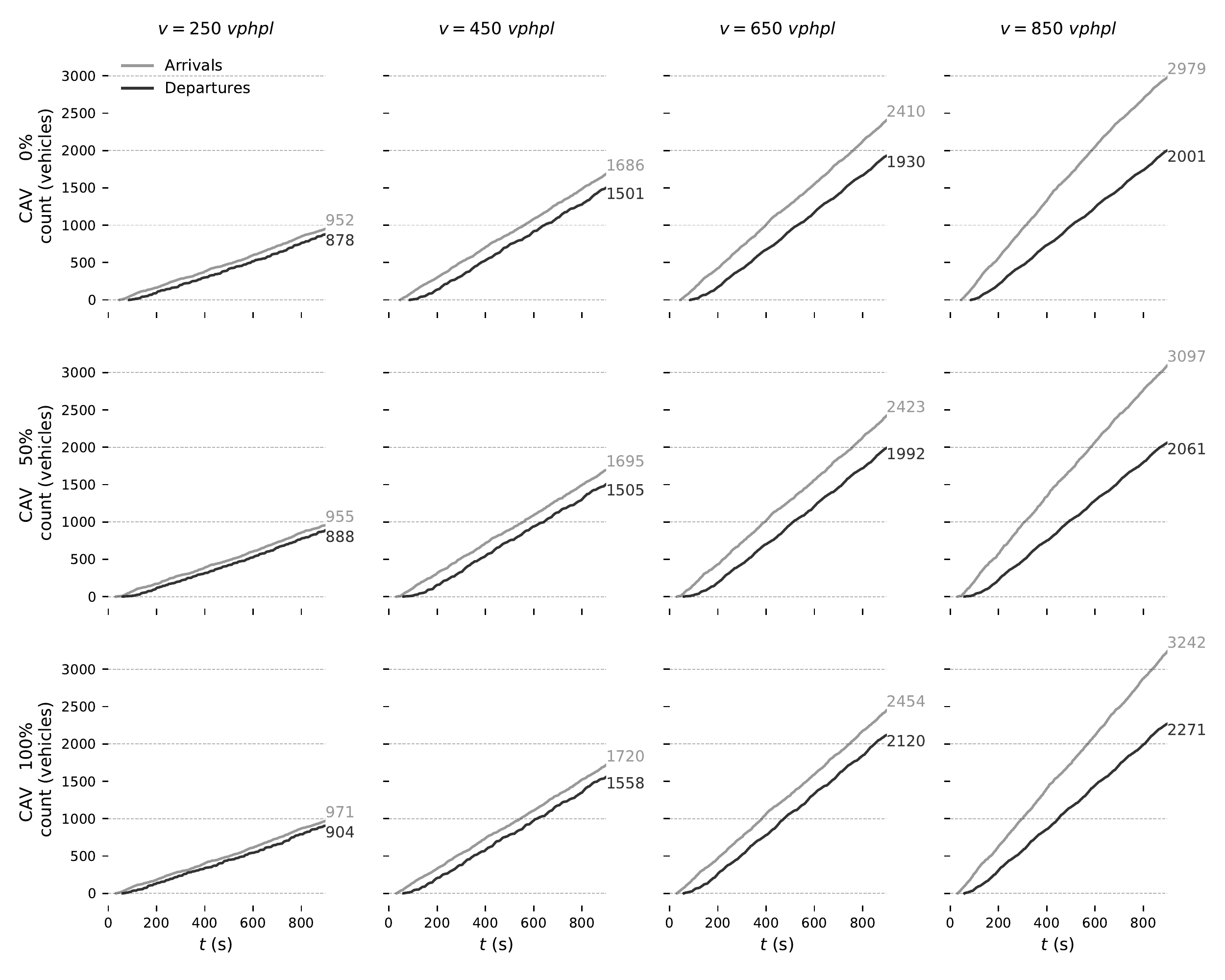}
  \caption{The cumulative arrival versus departure curves by SPaT optimization algorithm, traffic scenario, and CAV ratio.}\label{fig:cumulativeCurves}
\end{figure}

\begin{figure}[htbp]
  \centering
    \includegraphics[width=0.98\linewidth]{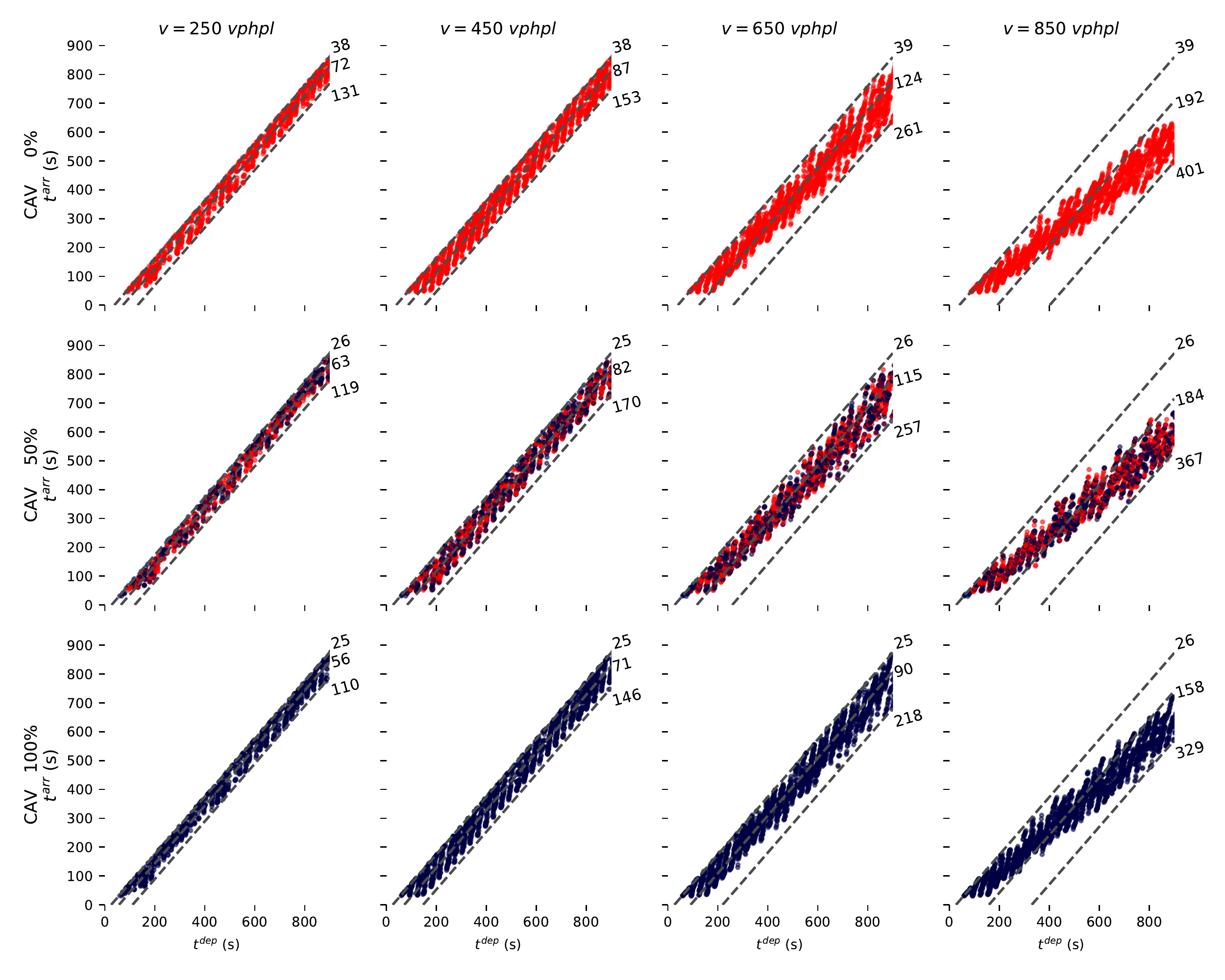}
  \caption{The departure time versus arrival time, $(t^{dep},t^{arr})$, of the vehicles. (The base lines in each panel indicate the minimum, average, and maximum travel times in \textit{seconds}.)}\label{fig:travelTimes}
\end{figure}

\begin{figure}[htbp]
  \centering
    \includegraphics[width=0.98\linewidth]{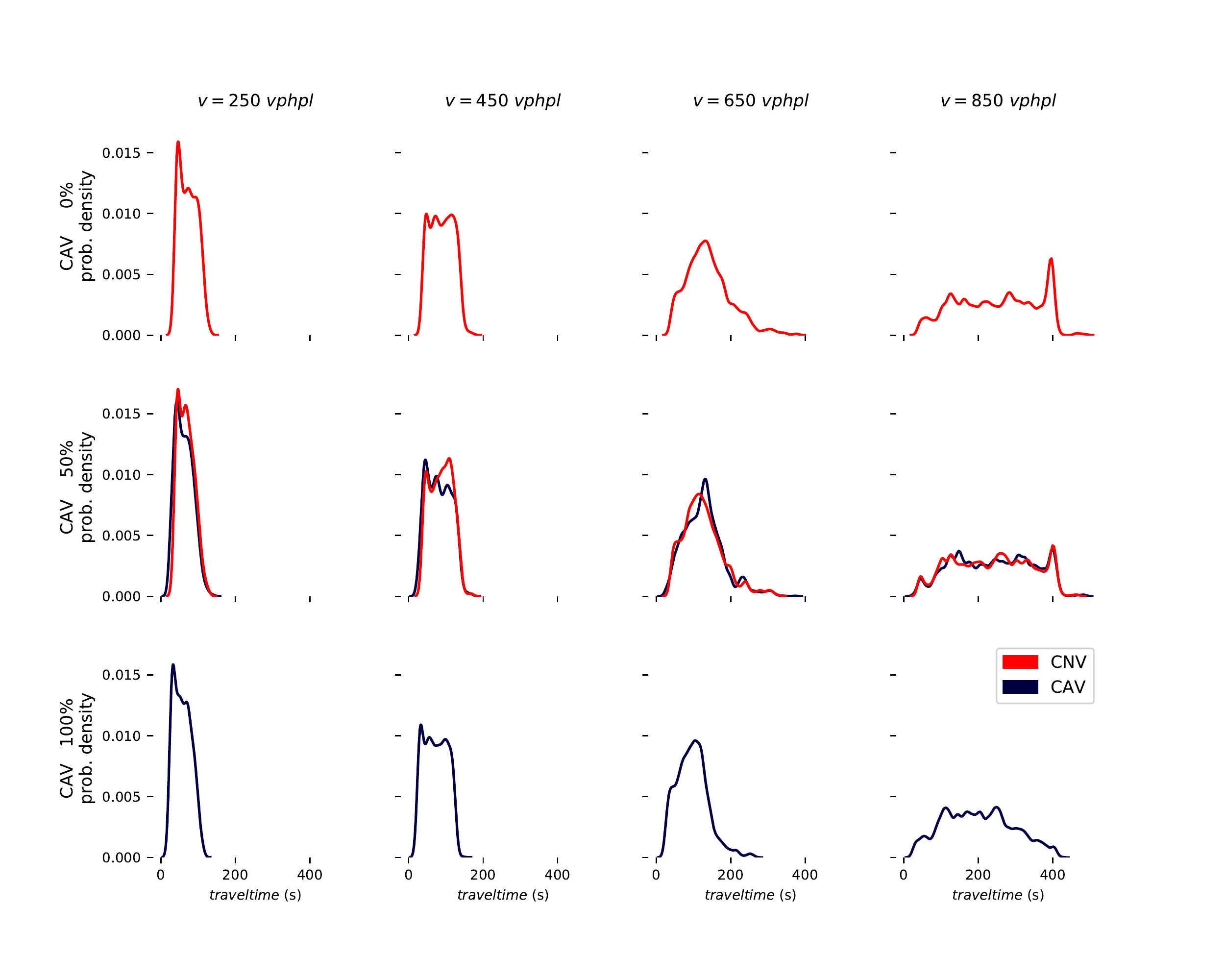}
  \caption{Travel time distribution per scenario and vehicle type.}\label{fig:travelTimesDist}
\end{figure}

We collected the following performance measures over the conducted simulations:
\begin{itemize}
    \item Average Throughput Rate
    is the average rate (in \textit{vphpl}) at which vehicles are served at the case study intersection.
    Table \ref{tab:dischRate} demonstrates the average throughput rates under each scenario's demand and CAV ratio.
    For a given CAV ratio (on the columns) and demand of vehicles per lanes (on the rows), the value is driven by averaging minutely throughput rate (in \textit{vphpl}) over the simulation period.
    
    \item Cumulative Arrival and Departure Curves
    which represent the count of vehicles at the communication distance and stop bar of all lanes, respectively.
    Fig.~\ref{fig:cumulativeCurves} plots cumulative arrival and departure curves at intersection level over 15 minutes of simulation. 
    
    \item Individual Vehicle Arrival and Departure Times
    measured at the communication distance and stop bar of all lanes, respectively.
    Fig.~\ref{fig:travelTimes} plots departure time versus arrival time of the vehicles.
    Every point in this figure represents a vehicle where its projections on the vertical and horizontal axis, i.e. $(t^{dep},t^{arr})$, correspond to the arrival and departure times, respectively.
    The three labeled dashed lines indicate the minimum, average, and maximum travel times (in \textit{seconds}) for each panel\textemdash equivalent to a scenario.
    
    \item Individual Vehicle Travel Times Probability Distribution
    measured over the simulation period for each scenario.
    Fig.~\ref{fig:travelTimesDist} illustrates the distribution of vehicle travel times (in \textit{seconds}) per vehicle type for all scenarios.
\end{itemize}

The simulation experiments resulted in the following findings:

\begin{itemize}
    \item As shown in Table \ref{tab:dischRate}, for under-saturated conditions, i.e. demand of 250 and 450 \textit{vphpl}, the vehicles discharged at the same rate they arrived.
    For comparison purposes, we derive an estimated throughput to compare the performance of the proposed algorithm under the range of demand and CAV ratios.
    Let's consider a pre-timed signal control algorithm that allocates green time to all feasible phases with an even probability.
    Under the best scenario, considering saturation flow of 1800 \textit{vphpl}, such an algorithm can serve vehicles at 535 \textit{vphpl} (computed by aggregating throughput rates in lanes of all phases with an equal probability weight of 1/8).
    We denote 535 \textit{vphpl} as the ideal throughput rate and use it as a base to measure the adaptiveness of the proposed algorithm.
    According to Table \ref{tab:dischRate}, for over-saturated conditions, the algorithm could reach a capacity ranging from 565 to 635 \textit{vphpl} (6 to 19\% higher than the throughput rate for the base case).

    \item According to Table \ref{tab:dischRate}, the CAV ratio has a significant impact on the throughput rate for capacity-bounded scenarios ($v=650,850$ \textit{vphpl}).
    For instance, at the demand of 850 \textit{vphpl}, the throughput rate increases by 12\% as the CAV ratio increases from 0 to 1.
    
    \item According to Fig.~\ref{fig:cumulativeCurves}, the throughput of the intersection increases by 13\% as the CAV ratio increases from 0 to 1.
    This is aligned with the throughput rate improvement.
    
    \item According to Fig.~\ref{fig:travelTimes}, the average travel time is lower by about 18 to 22\% in scenarios with full CAV traffic compared to those with full CNV traffic.
    \item According to Fig.~\ref{fig:travelTimes}, the maximum travel time decreases by about 16 to 18\% with CAV ratio ranging from 0 to 1.
    
    \item According to Fig.~\ref{fig:travelTimesDist}, the variance of travel times increases as the demand increases.
    For over-saturated conditions, the probability of higher travel times is significantly higher compared to under-saturated conditions.
    The mode travel time for the under-saturated conditions occurred at the low travel time side of the distribution.
    As the volume increases, the peak travel time moves toward the higher ranges.
    The highest travel time was reduced about 37\% when increasing the CAV ratio to 100\% with a volume of 850 \textit{vphpl}. 
\end{itemize}


\begin{table}[htbp]\footnotesize
	\caption{Average throughput rate (in \textit{vphpl}) per scenario.}\label{tab:dischRate}
	\centering
	\tiny
	\begin{tabularx}{1\linewidth} {Xcccccccccccc}
		\toprule
		& \multicolumn{11}{c}{CAV Percentage in Traffic} \\
		Demand (\textit{vphpl}) & 0\% & 10\% & 20\% & 30\% & 40\% & 50\% & 60\% & 70\% & 80\% & 90\%  & 100\%   \\
		\midrule
        250 & 248 & 249 & 248 & 248 & 248 & 248 & 248 & 248 & 247 & 248 & 248 \\
        450 & 447 & 447 & 447 & 447 & 444 & 444 & 444 & 445 & 446 & 445 & 446 \\
        650 & 566 & 569 & 573 & 577 & 581 & 586 & 593 & 600 & 608 & 617 & 626 \\
        850 & 568 & 572 & 576 & 580 & 584 & 589 & 594 & 603 & 613 & 622 & 635 \\
		\bottomrule
	\end{tabularx}
\end{table}

\section{Conclusions}\label{sec:conclusion}
This study developed an optimization-based intersection control algorithm which runs in real-time.
We formulate a mathematical model to jointly decide on SPaT and CAV trajectories in a mixed-traffic of CAVs and CNVs.
The model is reduced to a minimum cost flow network (MCF) optimization that provides SPaT and CAV departure times.
The MCF problem is designed with a realistic phasing scheme and is as low in complexity as a linear program, which makes it suitable for real-time applications. 
Trajectory planning problems are defined and solved for CAVs to optimize their path.
The overall framework is able to operate an intersection with over-saturated conditions through frequent re-optimization.\par

A case study was used to assess the effectiveness of the proposed algorithm.
The simulation results showed 6-19\% boost in throughput rate compared to the base case.
The intensity of operation enhancement, in terms of decrease in the mean and maximum of travel time distributions, increased with an increase in CAV ratio.
The findings support the hypothesis that the proposed algorithm enhances the intersection performance as CAV penetration rate and demand levels increases.\par

The enhanced intersection performance is achieved due to several considerations in the algorithm design.
The proposed model dynamically decides on SPaT and CAV trajectories based on an MCF model for SPaT and a linear program for CAV trajectory planning.
The proposed MCF-based SPaT model optimizes both the sequence and duration of phases to serve the highest number of lanes per selected phase.
Therefore, using the optimal SPaTs, the departure schedule of vehicles maximizes the green utilization at the intersection.
Also the proposed linear mathematical models minimizes the time-space that a follower or lead CAV uses to depart at the stop bar.
The framework frequently re-optimizes the decisions to flexibly sustain optimal performance level.\par

Several assumptions limit the scope of this study and left unanswered questions for future research.
This study did not consider the presence of pedestrians.
Similar to sensor technology for arriving vehicles, pedestrian and bicyclist arrivals can be used to extend the proposed algorithms to optimize operation for a broader group of users.\par

Our proposed model does not consider traffic signal preemption.
Also, per future technology advancements to increase range of DSRC and radio communication, it may become feasible to consider lane changing within the detection range.
The proposed RIO can be extended to advise CAVs to change lanes in a way that distributes them evenly among the lanes.
Also, the proposed algorithm should be scaled to coordinate operation at a network of intersections.\par

\section*{Acknowledgment}
This material is based upon work supported by the National Science Foundation (under Grant No. 1446813, titled: Traffic Signal Control with Connected and Autonomous Vehicles in the Traffic Stream).
The authors are grateful to the Econolite Group, Inc, and Emmanuel Posadas and Joe Crackel from the City of Gainesville Public Works Department for assistance with conducting field tests during the course of this research.
Any opinions, findings, and conclusions or recommendations expressed in this material are those of the authors and do not necessarily reflect the views of the National Science Foundation.
For more information on the project refer to \url{avian.essie.ufl.edu}.

\bibliographystyle{IEEEtran}
\bibliography{main}

\end{document}